\documentstyle[preprint,aps,floats,epsf]{revtex}
\begin{document}
\draft

\title{Droplet Formation in Quark-Gluon Plasma\\
       at Low Temperatures and High Densities}
\author{R.S. Bhalerao$^a$ and R.K. Bhaduri$^b$}
\address{  
$^a$ Department of Theoretical Physics, 
Tata Institute of Fundamental Research,\\
Homi Bhabha Road, Colaba, Mumbai 400 005, India\\
and\\
$^b$ Department of Physics, McMaster University, Hamilton, 
Ontario L8S 4M1, 
Canada}

\maketitle

\begin{abstract}
Considering the low-temperature ($T$) and high-baryon-number-density
($n_B$) region of the QCD phase diagram, we present a model for the
first-order phase transition between the quark-gluon plasma (QGP) and
the recently proposed colour superconducting phase. We study
nucleation of a droplet of the superconducting phase within the
metastable QGP gas. Numerical results for the activation energy,
radius and other physical parameters of the droplets, at various
temperatures, densities and gap parameters, are given. We have
estimated the latent heat of the phase transition. In the $T-n_B$
plane, we are able to demarcate the region of the superconducting
phase.
\end{abstract}



\vfill
\hrule
\vskip0.3cm
\noindent{E-mail: bhalerao@theory.tifr.res.in,
bhaduri@smiley.physics.mcmaster.ca}

\newpage
%--------------------------------------------------------------------
\section{Introduction}
%--------------------------------------------------------------------

Our understanding of the QCD matter at low temperatures ($T$) 
and high baryon number densities ($n_B$) 
has developed rapidly over the past few years;
for recent reviews, see \cite{raj00,sch00,alf01}. A likely
scenario is that in this region, quarks form Cooper pairs and new
condensates develop. One of the most interesting questions today
concerns the detailed mapping of the QCD phase diagram, especially in
the above region where QCD may exhibit a colour superconducting
phase. In this context, the light quark masses ($m_{u,d,s}$) are
expected to play an important role. Figure 1 shows two likely
scenarios corresponding to $m_s \gg m_{u,d} \neq 0$ and $m_s = m_{u,d}
\simeq 0$. The transition between the quark-gluon plasma (QGP) phase
and the colour superconducting phase is expected to be first order, in
QCD \cite{raj00,raj02}.

In this paper, we study the first-order phase transition between the
QGP phase and the colour superconducting phase (Fig. 1). Our approach
is phenomenological and is similar in spirit to that adopted by
C\^ot\'e and Kharchenko \cite{cot99}. They, however, studied the
Bose-Einstein condensation in a gas of trapped atomic hydrogen at
temperatures $\sim 50~ \mu K$ and densities $\sim 10^{14}$
cm$^{-3}$. Temperatures and densities that we shall consider here are,
of course, vastly different, and QGP has very little in common with
the atomic hydrogen gas. {\it Interestingly, however, as we shall see,
the same physics of activation energy barrier governs the formation of
droplets in both cases.}

It is important to get a handle on the cold and dense quark matter
because of its possible relevance to the neutron-star core. While the
region of {\it asymptotic} densities is amenable to rigorous
calculations, the same cannot be said about the region of {\it
intermediate} densities studied here. Both numerical (lattice) and
direct experimental studies of this region of the QCD phase diagram
are out of reach at the moment. Hence a phenomenological study such as
the present one is appropriate and worthwhile. It has allowed us to
make quantitative statements about the schematic phase diagram in
Fig. 1.

Alford et al. \cite{alf98} have discussed droplet formation in the
transition between the hadronic matter phase and the 2SC phase
(Fig. 1), at $T
= 0$, as a function of density. The chiral symmetry is restored inside
their droplets (which they call nucleons) but is broken outside.
Neergaard and Madsen \cite{nee00} also have studied the droplet
formation but in the QGP to hadronic matter transition occurring at
low baryon number densities. 
In this paper, we have considered the QGP to colour-superconducting
phase transition occurring
at high $n_B$. {\it To our knowledge, droplet formation in this
transition has not been discussed previously in the literature.}
Our droplets are the opposite of those in \cite{alf98} in the sense
that the chiral symmetry is broken inside our droplet, but is restored
outside.

%--------------------------------------------------------------------
\section{Model}
%--------------------------------------------------------------------

Consider a gas of weakly interacting quarks, antiquarks and gluons in
a volume $V$. We take $q$ and $\bar q$ to be the massless current
quarks (Fig. 1). We are interested in low temperatures ($T$) and high
number densities ($n$) such that $n \lambda^3 \gg 1$ where $\lambda$
is the thermal wave length (see Appendix). We also need the baryon
number density $n_B = (n_q - n_{\bar q})/3$ to be large compared to
that for the normal nuclear matter, namely 0.17 fm$^{-3}$.

If the interactions among the particles are treated to the lowest
order in $\alpha_s$, the strong coupling constant (Fig. 2), the baryon
number density is given by ($c=\hbar=1$)
\begin{equation}
n_B = \left( 1 - \frac {2 \alpha_s} {\pi} \right)
\frac{d_q}{3} \left[ \frac{\mu_q (kT)^2}{6} + 
\frac{\mu_q^3}{6 \pi^2} \right],
\end{equation}
where $d_q =$ 2~(spin) $\times$ 3~(colour) $\times$ 3~(flavour) = 18
is the quark degeneracy factor, $k$ is the Boltzmann constant and 
$\mu_q$ is
the quark chemical potential \cite{mul85}. For a given baryon number
density $n_B$ and temperature $T$, Eq. (1) uniquely fixes
$\mu_q$. This in turn determines the total energy and pressure of the
gas as
\begin{eqnarray}
E_{gas} &=& 
\left( 1 - \frac {15 \alpha_s} {4 \pi} \right)
 d_gV \frac{\pi^2 (kT)^4}{30}  \nonumber \\
 &+&
 d_qV \left[ \left( 1 - \frac {50 \alpha_s} {21 \pi}
 \right)
 \frac{7 \pi^2 (kT)^4}{120} +
 \left( 1 - \frac {2 \alpha_s} {\pi} \right) \left( 
 \frac{\mu_q^2 (kT)^2}{4} + 
  \frac{\mu_q^4}{8 \pi^2} \right) \right],\\
P_{gas} &=& E_{gas}/3V.
\end{eqnarray}
Here $d_g =$ 2~(spin) $\times$ 8~(colour) = 16 is the gluon degeneracy
factor. It was possible to obtain the above analytic expressions
because we have considered the gas of quarks as well as antiquarks,
rather than a gas of quarks alone \cite{mul85}. At low $T$ and high
$n_B$, the $\mu_q$ terms in the pressure dominate and the gluonic
pressure arising from the first term in Eq. (2), is quite negligible.

We now consider the first-order phase transition from QGP to the
superconducting phase as the temperature is lowered (Fig. 1). In the
latter phase, as a result of the attractive one-gluon-exchange
interaction between quarks in the colour ${\bf \bar 3}$ channel,
quarks near the Fermi surface tend to pair up as bosons. These
are the Cooper pairs which we assume to be massless. Density
fluctuations in the metastable QGP gas may provide suitable conditions
for nucleation and growth of a droplet of the superconducting phase.
The high-temperature phase (QGP) is present around the droplet and the
low-temperature superconducting 
phase exists inside the droplet. We will
presently explore the energetics for the formation of such a droplet
by calculating the energy barrier between the two phases.

We have given above the expressions for number density, energy and
pressure of the QGP; see Eqs. (1)-(3). We now derive corresponding
expressions for the superconducting phase inside the droplet. Before
we begin, it is necessary to recall important differences between the
BCS superconductor and the colour superconductor.
(a) In the BCS theory of superconductivity the ratio of the energy gap
($\Delta$) to the Fermi kinetic energy for a typical metal is of the
order of $10^{-4}$. In contrast, in the problem under consideration,
the gap $\Delta$ is estimated to vary between several tens of MeV to
about 100 MeV \cite{raj00,sch00,alf01}, and the quark chemical
potential $\mu_q$ appearing in Eq. (1) takes values between $\sim 250$
and $\sim 600$ MeV for the $n_B$ and $T$ considered here.
(b) In the BCS superconductor, the Debye energy $\omega_D$ is much
smaller than the chemical potential $\mu$, while in the colour
superconductor, the two may be of the same order of magnitude.
(c) Finally, if $n$ is the density of electrons and $\xi$ the
Pippard coherence length which measures the spatial extension of the
Cooper pair wave function, then in the BCS theory, $n\xi^3 \gg 1$. In
the present case, on the other hand, $\xi \simeq 1/(\pi \Delta) \simeq
0.6$ fm for $\Delta = 100$ MeV, and $n\xi^3 < 1$
for the densities considered in
this paper. This suggests that the $qq$ pairs are rather compact
objects unlike Cooper pairs of electrons in a metal. For a good
discussion of these and related issues, see a review article by
Kerbikov \cite{ker01}.
We shall make use of the above points while deriving expressions for
thermodynamics quantities of the droplet. For the sake of simplicity,
we shall assume that all quarks in the energy interval
$(\mu_q-\Delta,\mu_q)$ form Cooper pairs and all other quarks remain
unpaired, and shall ignore other interactions.

Consider a spherical region of radius $R_d$, volume $V_d$, containing
$N_1$ quarks out of which $N_2$ remain unpaired and $N_3$ undergo
Cooper pairing, when the pairing interaction is ``switched on''. We
have
\begin{eqnarray}
N_1 &=& \frac{d_q V_d}{(2 \pi)^3} \int^\infty_0 4 \pi p^2 f dp,\\
N_2 &=& N_1 - \frac{d_q V_d}{(2 \pi)^3} \int^{\mu_q}_{\mu_q-\Delta} 
4 \pi p^2 f dp,\\
N_3 &=& N_1 - N_2,
\end{eqnarray}
where $f=\left [ \exp ~\beta(p-\mu_q)+1 \right ]^{-1}$ with 
$\beta=1/(kT)$.
The corresponding energies $E_{1,2,3}$ are
\begin{eqnarray}
E_1 &=& \frac{d_q V_d}{(2 \pi)^3} \int^\infty_0 4 \pi p^3 f dp,\\
E_2 &=& E_1 - \frac{d_q V_d}{(2 \pi)^3} \int^{\mu_q}_{\mu_q-\Delta} 
4 \pi p^3 f dp,\\
E_3 &=& \left( \frac{N_3}{2} \right)
\left( \frac{\pi}{R_d} - \Delta \right).
\end{eqnarray}
In the last equation $(N_3/2)$ is the number of Cooper pairs
(bosons) in the droplet and $(\pi /R_d)$ is the energy of the
ground state of the single-particle spectrum of a spherical cavity of
radius $R_d$.
Due to the small size of the droplet, even the lowest excited state
has energy which is very large compared to the temperature. Therefore
all the $qq$ pairs tend to occupy the ground state \cite{cot99}.
Next, entropies $S_{1,2,3}$ are given by
\begin{eqnarray}
S_1 &=& -\frac{d_q V_d}{(2 \pi)^3} \int^\infty_0 4 \pi p^2 dp
\left[ f \ln f + (1-f) \ln (1-f) \right],\\
S_2 &=& S_1 + \frac{d_q V_d}{(2 \pi)^3} \int^{\mu_q}_{\mu_q-\Delta}
4 \pi p^2 dp \left[ f \ln f + (1-f) \ln (1-f) \right],\\
S_3 &=& 0.
\end{eqnarray}
$S_3$ can be taken to be zero for the reasons given above.
Finally, the pressures $P_{1,2,3}$ are calculated as follows:
\begin{equation}
P_1 = E_1 /(3V_d).
\end{equation}
The thermodynamic potential $\Omega$ for massless quarks is given by
\[
\Omega = -kT \sum \ln \left [ 1 + \exp \beta (\mu_q - p) \right ].
\]
Hence
\begin{eqnarray}
P_2 &=& - \frac{\Omega}{V_d} = 
\frac{kTd_q}{(2 \pi)^3}\left(\int_0^\infty-\int_{\mu_q-\Delta}
^{\mu_q}\right)4 \pi p^2 dp \ln\left[1 + \exp \beta (\mu_q - p) \right ]
\nonumber \\
&=&\frac{E_1}{3V_d}-\frac{kTd_q}{2 \pi^2}\int_{\mu_q-\Delta}
^{\mu_q} p^2 dp \ln\left[1 + \exp \beta (\mu_q - p) \right ]. \nonumber
\end{eqnarray}
Performing the last integration by parts and rearranging the
terms, one finally gets
\begin{equation}
P_2=\frac{E_2}{3V_d}-\frac{kTd_q}{6 \pi^2} \{
\mu_q^3 \ln 2 - (\mu_q-\Delta)^3 \ln \left [ 1+\exp(\beta \Delta) 
\right] \}.
\end{equation}
The quantal pressure $P_3$ is given by
\begin{equation}
P_3 = - (\partial E_3/\partial V_d)_{N_3}
= N_3 /8 R_d^4.
\end{equation}
It is easy to verify that if $\Delta$ is set equal to zero, i.e. if no
pairing is allowed to take place, $N_3$, $E_3$, $P_3$ vanish and
$N_2$, $E_2$, $S_2$, $P_2$ become equal to $N_1$, $E_1$, $S_1$, $P_1$,
respectively.
Contribution of antiquarks to the above thermodynamic quantities is
smaller by many orders of magnitude. Although we have included it in
our calculations its effect is negligible.

The droplet pressure $P_d$ is given by
\begin{equation}
P_d=P_2+P_3+\bar P_2+\bar P_3+P_{gluons},
\end{equation}
where $P_2$ and $P_3$ are as in Eqs. (14)-(15), $\bar P_2$ and $\bar
P_3$ are the corresponding pressures due to antiquarks and
$P_{gluons}$ is the pressure due to gluons in the droplet. 

Consider once again the spherical region of radius $R_d$. Initially,
i.e. before the pairing interaction is switched on, there are $N_1$
quarks, $\bar N_1$ antiquarks and a certain number of gluons in it. In
the final state, i.e. when the pairing interaction is on, there are
$N_2$ quarks, $(N_3/2)$ quark pairs, $\bar N_2$ antiquarks, $(\bar
N_3/2)$ antiquark pairs and the same number of gluons as before. The
activation energy ($A$) is defined as the difference between the
Helmholtz free energies ($F$) in the final and initial states. Since
the gluonic contribution cancels out, one gets
\begin{eqnarray}
A &=& F(N_2)+F(N_3/2)-F(N_1)+F(\bar N_2)+F(\bar N_3/2)-F(\bar N_1)
\nonumber \\
&=& E_2-TS_2+E_3-E_1+TS_1+\bar E_2-T\bar S_2+\bar E_3-\bar E_1+T
\bar S_1.
\end{eqnarray}
In the limit $\Delta \rightarrow 0$, this $A$ vanishes. 

The reader may wonder why surface tension of the droplet makes no
appearance in our formalism. In the classical theory of nucleation,
there are two competing contributions to the energy of a droplet:
droplet formation reduces bulk free energy but only 
at the cost of surface
energy. The former contribution wins if the droplet radius exceeds a
certain critical value, while the surface tension ensures that a
droplet with a smaller radius shrinks and disappears. In the formalism
presented here, energy of the droplet is calculated quantum
mechanically. It is clear from Eq. (9) that small droplets have large
quantal energy and tend to dissolve back into the QGP gas. This is
similar to the effect of the surface tension in the classical theory.
Moreover, the thermal wavelength ($\lambda$)
of the quarks is so large that a
classification of particles as those localized on the surface and
those in the interior breaks down. In short, we did not neglect the
surface energy; it is included in the total free energy of the quantal
droplet.

%--------------------------------------------------------------------
\section{Results and Discussion}
%--------------------------------------------------------------------

The precise boundaries of the various phases in Fig. 1 are not
known. Since the baryon number density ($n_B$) of the normal nuclear
matter is $\sim 0.17$ fm$^{-3}$, we have performed numerical
calculations for $n_B=$ 0.5 to 3 fm$^{-3}$ and temperature $T=$ 0 to
100 MeV. The resulting quark chemical potential $\mu_q$ varies over
$\sim 250 - 600$ MeV and the baryon chemical potential\footnote{This
definition is applicable only for a gas of quarks, and not for the
normal nuclear matter.} $\mu_B \equiv 3 \mu_q$ varies over $\sim 750 -
1800$ MeV. This fully covers the densities likely to be seen in the
neutron-star cores. We have considered four different values of the
superconducting gap parameter $\Delta$, namely, 1, 10, 50 and 100 MeV.

Since we are interested in a gas
of weakly interacting quarks, antiquarks and gluons, we take $\alpha_s$
to be zero. Small values of $\alpha_s$ do not affect the results in
Eqs. (1)-(3) significantly. 
We first choose a set of values for $n_B$ and $T$, and solve Eq. (1)
numerically to determine $\mu_q$. This
$\mu_q$ is then used to evaluate $P_{gas}$ as given by Eq. (3). Also,
once $\mu_q$ is determined, $n_q$ and $n_{\bar q}$ can be evaluated
separately by using
\begin{eqnarray}
n_q&=&\frac{d_q}{(2 \pi)^3} \int_0^\infty 4 \pi p^2 f dp,\nonumber \\
n_{\bar q}&=& n_q(\mu_q \rightarrow -\mu_q) \nonumber
\end{eqnarray}
(Above integrations cannot be done analytically, and so it is not
possible to get closed expressions like Eq. (1), for $n_q$ and $
n_{\bar q}$.) Next, we choose a value for $\Delta$, and treating the
droplet radius $R_d$ as a free parameter, calculate various other
quantities; see Eqs. (4)-(17). Some of these results are shown in
Figs. 3-4. 

In our formalism, as in the classical theory, the
activation energy when plotted as a function of the radius (Fig. 3),
initially increases from zero, attains a maximum, then starts
decreasing monotonically eventually becoming negative for a
sufficiently large value of the radius. In other words, the activation
energy exhibits a barrier.
{\it The barrier originates as follows}. Consider Eq. (17). As already
stated, the antiquark terms are extremely small. The terms $TS_1$ and
$TS_2$ are also negligible and in fact they tend to cancel each other.
That leaves three terms, namely $E_1,~E_2$ and $E_3$. For small radii,
$E_2$ and the quantal energy term $E_3$ dominate pushing the
activation energy curve upwards. For large radii, the $E_1$ term
dominates pulling the curve to the negative domain.
As expected, the barrier becomes wider if
$T$ is raised or if $n_B$ is lowered.  The critical radii can be read
off from Fig. 3. Droplets with larger radii readily grow because that
tends to reduce the activation energy. This in turn easily brings
about the transition from QGP to the superconducting phase. 

Figure 4 shows the ratio of the droplet pressure to the pressure of
the surrounding gas, for $\Delta=100$ MeV. This ratio is significantly
larger than unity for small droplets, which will tend to push the
radii to larger values, thereby helping to bring about the QGP to
superconducting phase transition. In the limit $\Delta \rightarrow 0$,
this ratio tends to unity and is thus incapable of pushing the droplet
radii to larger values.

It is interesting to know the properties of the droplet when the
pressure inside the droplet (Eq. 16) equals the pressure of the
surrounding QGP gas (Eq. 3). These results are shown in Figs. 5-8.
The condition $P_d = P_{gas}$ fixes $R_d$. We solve the equation $P_d
= P_{gas}$ numerically to determine $R_d$, for various values of
$n_B$, $T$ and $\Delta$. $n_B$ was varied from 0.5 to 3 fm$^{-3}$ in
the steps of 0.5 fm$^{-3}$.
Figure 5 shows the results for $R_d$. As $T$ and/or $n_B$ are
increased, $R_d$ decreases. However, $R_d$ is found to be rather
insensitive to the variation in $n_B$. Although we have varied $T$
between 0 and 100 MeV, temperatures relevant for neutron stars are
close to zero. Because of the small value of $R_d$, the energy of even
the lowest excited state inside the droplet turns out to be several
hundreds of MeV which is large compared to the temperature,
effectively forcing all Cooper pairs (bosons) to occupy the ground
state. {\it Note also that at low temperatures, the droplet radius
$R_d$ is larger than the Pippard coherence length $\xi \simeq 1/(\pi
\Delta)$.}

Figure 6 shows the activation energy $A$. First see the upper half of
Fig. 6. For a fixed $n_B$, as $T$ increases, $A$ changes its sign from
negative to positive. This is expected because, at higher $T$, droplet
formation is energetically unfavourable.  Secondly, as $n_B$ increases
(see curves labelled {\sl a} to {\sl f} ), $A$ remains negative up to
higher and higher $T$. This is also expected because at higher $n_B$,
the transition to the superconducting phase is facilitated. 
Now see the lower half of Fig. 6. It shows the $A=0$ contour, in the
$T-n_B$ plane. This is also the phase transition curve separating the
QGP phase at high $T$ and the superconducting phase at low $T$: Recall
$A \equiv F_{SC}-F_{QGP}$ where $F$ are the free energies of the two
phases. Both $F_{SC}$ and $F_{QGP}$ decrease as $T$ increases and
cross each other at some temperature, say $T_c$ (Fig. 7). Since the
system follows the path of minimum $F$, there is a kink in $F(T)$. In
other words, the slope $dF/dT=-S$ has a discontinuity, indicating a
first-order phase transition at $T_c$. Thus the $A=0$ contour serves
to demarcate the region of the superconducting phase. Note that as
$n_B$ increases, the $A<0$ region in Fig. 6 survives up to higher and
higher $T$.

We now present an estimate of the latent heat ($L$) associated with
the first-order phase transition considered in this paper. The
transition temperature $T_c$ for any given $n_B$ can be determined
from Fig. 6. The latent heat is then estimated as $L=T_c \times
(S_1(T_c)-S_2(T_c))$. For $\Delta=100$ MeV, this turns out to be $\sim
150$ to $\sim 250$ MeV and it increases with $n_B$. In the limit
$\Delta \rightarrow 0$, the two free-energy curves in Fig. 7 coincide,
$A$ vanishes, $S1=S2$, there is no discontinuity in entropy, and the
latent heat is zero. There is only one phase namely QGP, and there is
no question of any phase transition.

Finally, Fig. 8 shows 3-dimensional surface plots of the (normalized)
activation energy $A'$ defined as $A' \equiv
(F_{SC}-F_{QGP})/(F_{SC}+F_{QGP})$ as a function of $n_B$ and $T$, for
$\Delta=1$ and 100 MeV. It is clear that in the limit $\Delta
\rightarrow 0$, the activation energy surface is flat and there is no
energy advantage for the QGP to superconducting phase transition.
Indeed there would be only one phase, namely QGP, which would extend
up to the $T=0$ axis. On the other hand, for $\Delta=100$ MeV, the
surface is such that the system would like to ``roll down'' the slope
thereby bringing about the phase transition.

It is interesting to compare the activation energy results for the QCD
phase transition, obtained here with those for the Bose-Einstein
condensation in a gas of trapped atomic hydrogen, presented in
\cite{cot99}. In \cite{cot99} two gas densities were considered:
$10^{14}$ cm$^{-3}$ and $7 \times 10^{13}$ cm$^{-3}$ both at the same
temperature $50~ \mu K$. We have repeated their calculations for
$T=45-55 ~\mu K$. 
Their activation energy exhibits a barrier, and they denote the
maximum value of the activation energy by $A_0$.
Consider the ratio $A_0/kT$ at the peak position and
the corresponding number $\tilde N_0$ of particles in the droplet. In
\cite{cot99} these two quantities are highly sensitive to the changes
in the gas density and temperature.\footnote{There is a misprint in
\cite{cot99}: in Fig. 1, the y-axis label should read $A(10^{-10})$
a.u. For the inset, it is $A(10^{-8})$ a.u.} For instance, for the
above tiny change in the gas density, the ratio $A_0/kT$ changes from
4.8 to 104.1, and $\tilde N_0$ changes from 18 to 3862. Similar
sensitivity was seen when $T$ was changed by a few $\mu K$. In
contrast, the results presented here are relatively stable with
respect to changes in $n_B$ and $T$.

In conclusion, we have studied hitherto unexplored first-order phase
transition from QGP to a colour-superconducting phase, and have
presented a simple model of the transition with very few
assumptions. It provides numerical support to the prevalent ideas of
phase structure of QCD in the low-temperature,
high-baryon-number-density regime. We have studied properties of the
droplets of the superconducting phase, formed at various points in the
$T-n_B$ plane, for several values of the gap parameter. 
We have been able to estimate the extent of the
superconducting phase in this plane and the latent heat of the
transition. At low temperatures, droplet radii are found to be larger
than the Pippard coherence length which measures the spatial extent of
the $qq$ Cooper pairs.

%----------------------------------------------------------------------
%Acknowledgments
%----------------------------------------------------------------------

%\newpage
\bigskip\bigskip\bigskip

RSB acknowledges the hospitality of the McMaster University where this
work was initiated. We thank Mark Alford and Mustansir Barma for their
comments on this work. 
This research was partially supported by a grant from NSERC
(Canada).

%---------------------------------------------------------------------
%      Appendix
%---------------------------------------------------------------------

\newpage
\bigskip\bigskip\bigskip
\begin{center}
{\bf Appendix}
\end{center}

If $f(p)$ is the distribution function of a gas in equilibrium, the
number density $n$ is given by $n=\int f(p) d^3p/(2 \pi \hbar)^3$. 
For a nonrelativistic classical gas at temperature $T$, one has
\[
f = \exp \left[ -\beta \left( \frac{p^2}{2m} - \mu \right) \right],
\]
where $\beta=1/kT$ and $\mu$ is the chemical potential. Hence
\[
n = \exp(\beta \mu) (2 \pi \hbar^2 / m k T)^{-3/2} \equiv \exp(\beta
\mu)
\lambda^{-3},
\]
where $\lambda$ is the thermal wave length.

For an extreme relativistic gas of massless classical particles,
\[
f = \exp \left[- \beta (pc - \mu) \right],
\]
and hence
\[
n = \exp(\beta \mu) (c \hbar \pi^{2/3} / k T)^{-3}.
\]
We define the thermal wave length in this case as
$
\lambda = c \hbar \pi^{2/3} / k T,
$
so that the fugacity $\exp(\beta \mu)$ once again equals $n \lambda^3$.

%----------------------------------------------------------------------
% References
%----------------------------------------------------------------------

%---------------------------------------------------------------------
%      Figure Captions
%---------------------------------------------------------------------

\newpage
\begin{center}
FIGURE CAPTIONS
\end{center}

FIG. 1.
Schematic phase diagrams of QCD: $T$ is the temperature and $\mu_B$ is
the baryon chemical potential. $E$ is the critical point at which the
line of first-order phase transitions ends. $2SC$ and $CFL$
(Colour-Flavour-Locked) are 2-flavour-like and 3-flavour-like colour
superconducting phases, respectively. The dashed line denotes the
critical temperature at which quark-quark pairing vanishes. The thick
arrow indicates the first-order phase transition considered in this
paper. For the normal nuclear matter at $T=0$, $\mu_B$ should be
defined as $m_N + \hbar^2 k_F^2 /2 m_N$ where $k_F \simeq 1.4$
fm$^{-1}$ 
\cite{bha75}
and $m_N$ is the nucleon mass, which gives $\mu_B \simeq 980$ MeV.
(Figure adapted from \cite{raj00}.)

%---------------------------------------------------------------------

FIG. 2.
Feynman diagrams contributing to the equation of state of
quark-gluon plasma in order $\alpha_s$. Curly lines: gluons, dashed
lines: ghost particles, solid lines: quarks.

%---------------------------------------------------------------------

FIG. 3.
Activation energy $A$ vs droplet radius $R_d$ for various gap
values $\Delta$. Baryon number density $n_B$ (fm$^{-3}$) and
temperature $T$ (MeV) are: (a) 0.5 and 5 (b) 0.5 and 100 (c) 3 and 5
(d) 3 and 100, respectively.

%---------------------------------------------------------------------

FIG. 4.
Ratio of the pressure of the droplet to the pressure of the
surrounding QGP gas vs droplet radius $R_d$, at gap $\Delta=100$
MeV. Curve labels are as in Fig. 3.

%---------------------------------------------------------------------

FIG. 5.
Droplet radius $R_d$ vs temperature $T$ for various gap
values $\Delta$. Results are insensitive to $n_B$, so not all curves
are labelled.

%---------------------------------------------------------------------

FIG. 6.  
Upper panel: 
Activation energy $A$ vs temperature $T$.
Baryon number densities $n_B$ (fm$^{-3}$) are (a) 0.5 (b) 1 (c)
1.5 (d) 2 (e) 2.5 (f) 3.
Lower panel:
The $A=0$ contour or the phase transition line in the $T-n_B$ plane.
SC: superconducting phase.

%---------------------------------------------------------------------

FIG. 7.  
Free energies $F_{QGP}$ and $F_{SC}$ in the QGP and superconducting
phases, respectively, vs temperature $T$.

%---------------------------------------------------------------------

FIG. 8.  
Three-dimensional surface plots of the activation energy $A'$ for
$\Delta=1$ MeV (upper part) and $\Delta=100$ MeV (lower part),
as a function of $n_B$ (fm$^{-3}$) and $T$ (MeV).

%---------------------------------------------------------------------

\newpage
\pagestyle{empty}
\begin{figure}
\begin{center}
\epsfxsize=9.5in
\epsfbox{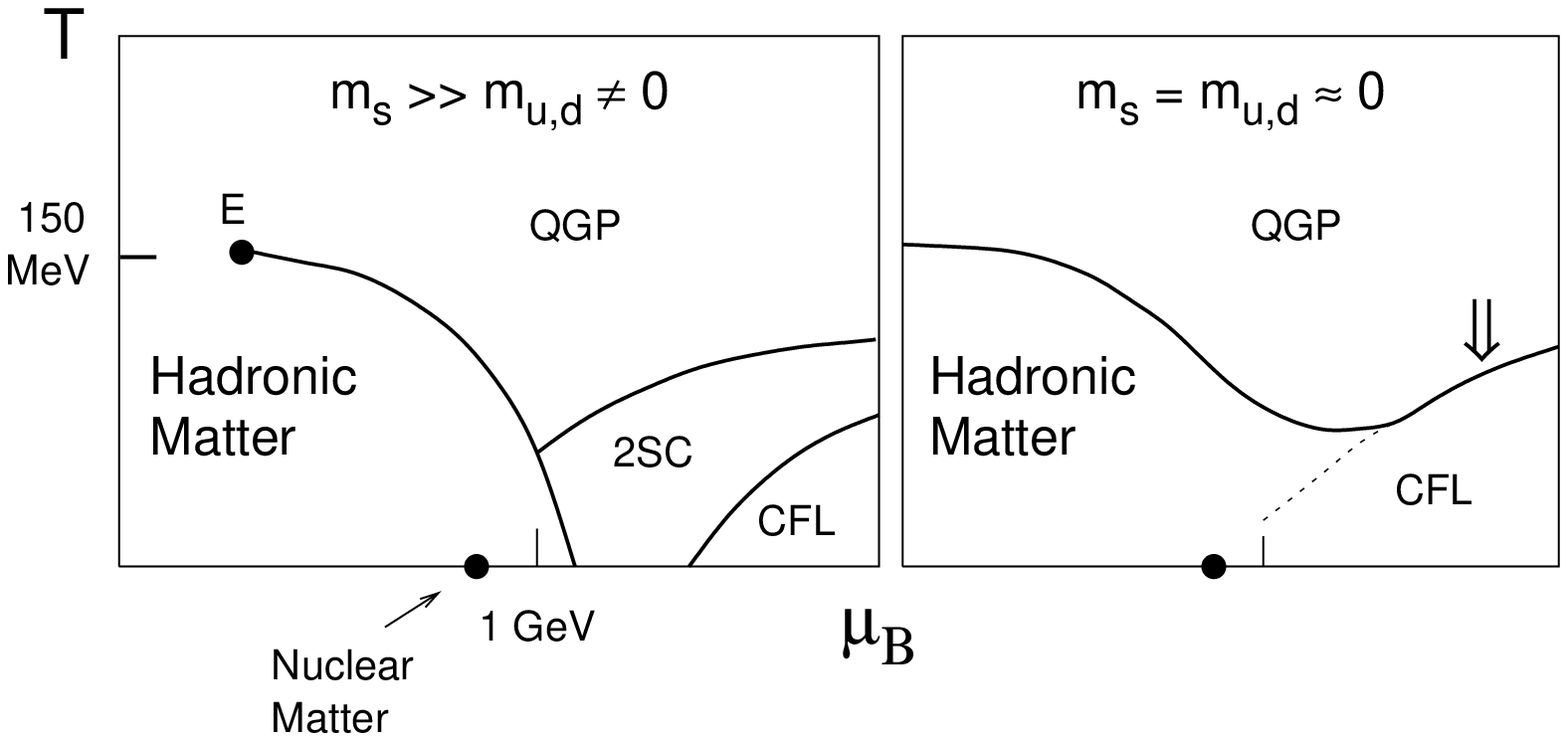}
\caption{}
\end{center}
\end{figure}

\newpage
\begin{figure}
\begin{center}
\epsfxsize=8.3in
\epsfbox{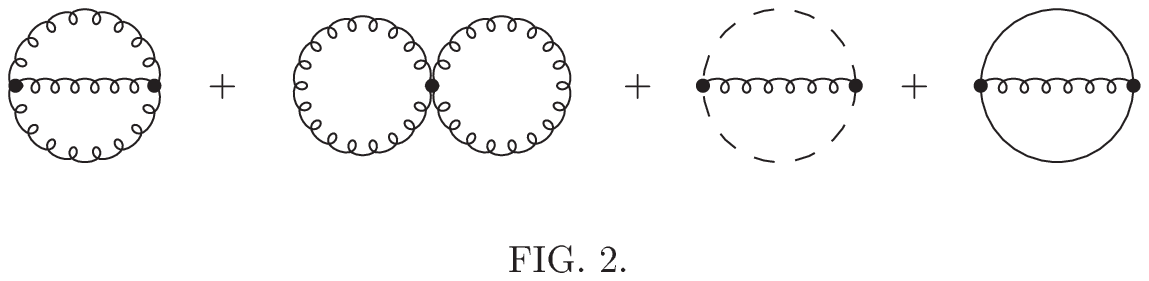}
\caption{FIG. 2}
\end{center}
\end{figure}

\newpage
\begin{figure}
\begin{center}
\epsfxsize=8.5in
\epsfbox{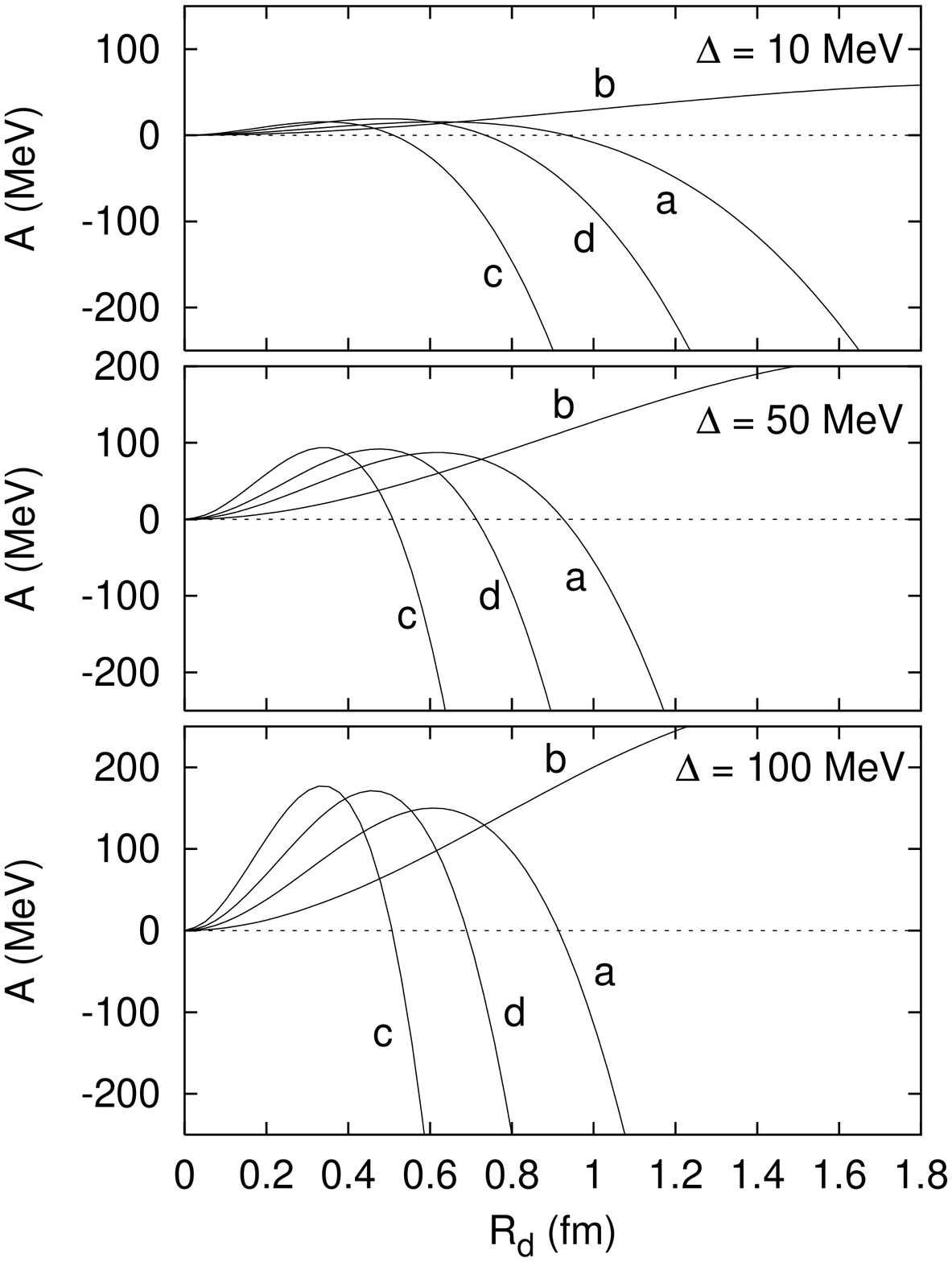}
\caption{}
\end{center}
\end{figure}

\newpage
\begin{figure}
\begin{center}
\epsfxsize=6.7in
\epsfbox{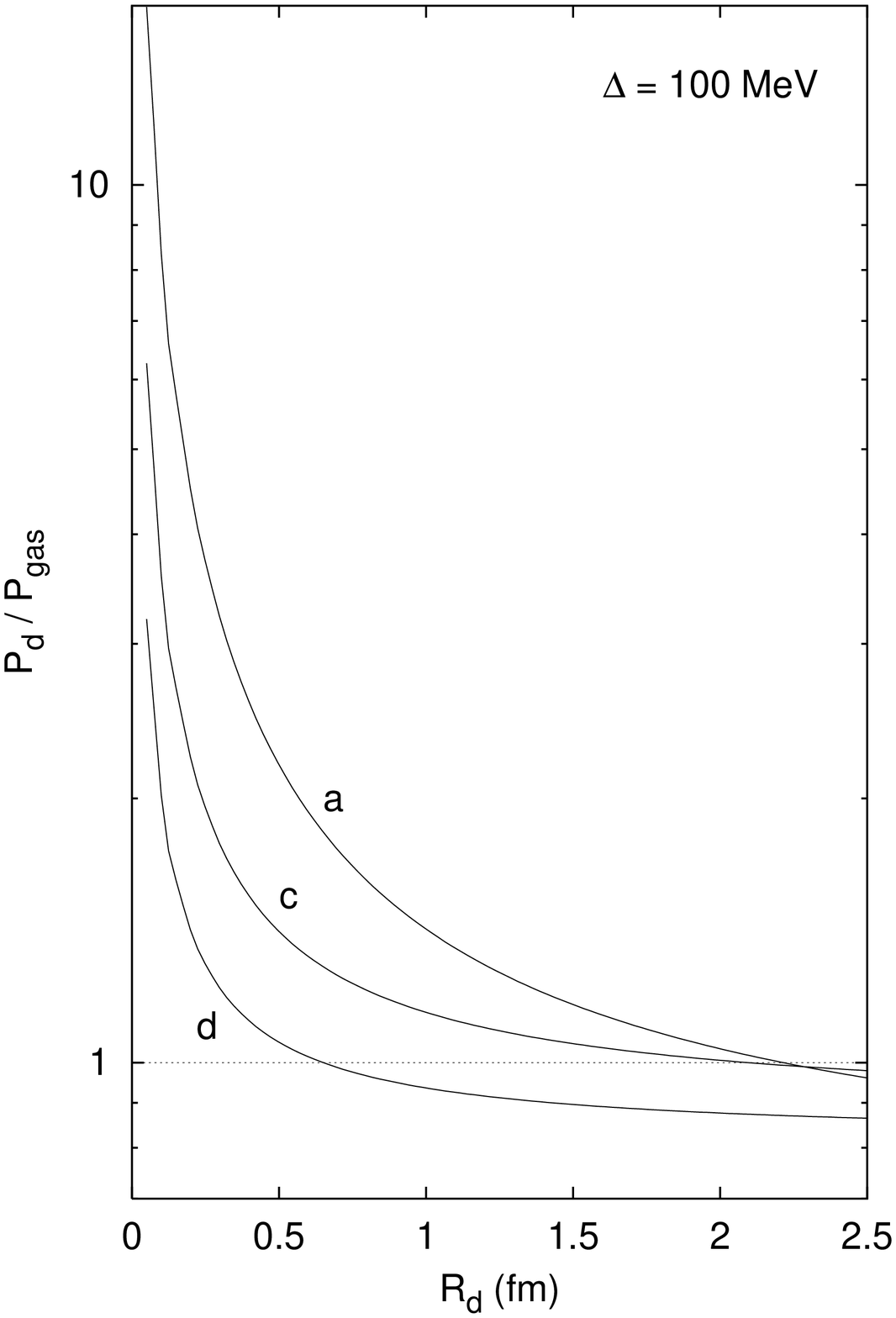}
\caption{}
\end{center}
\end{figure}

\newpage
\begin{figure}
\begin{center}
\epsfxsize=8.5in
\epsfbox{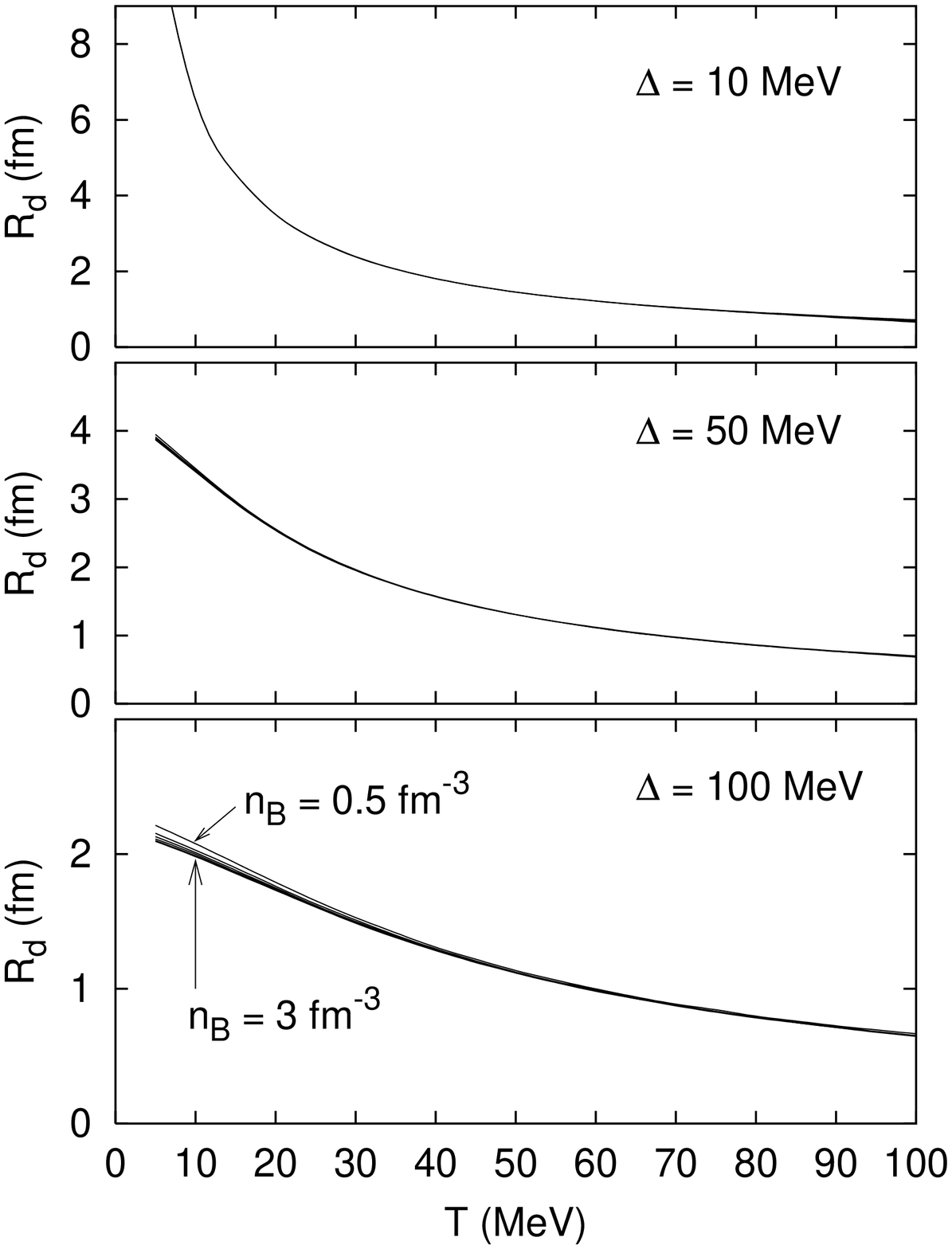}
\caption{}
\end{center}
\end{figure}

\newpage
\begin{figure}
\begin{center}
\epsfxsize=8.5in
\epsfbox{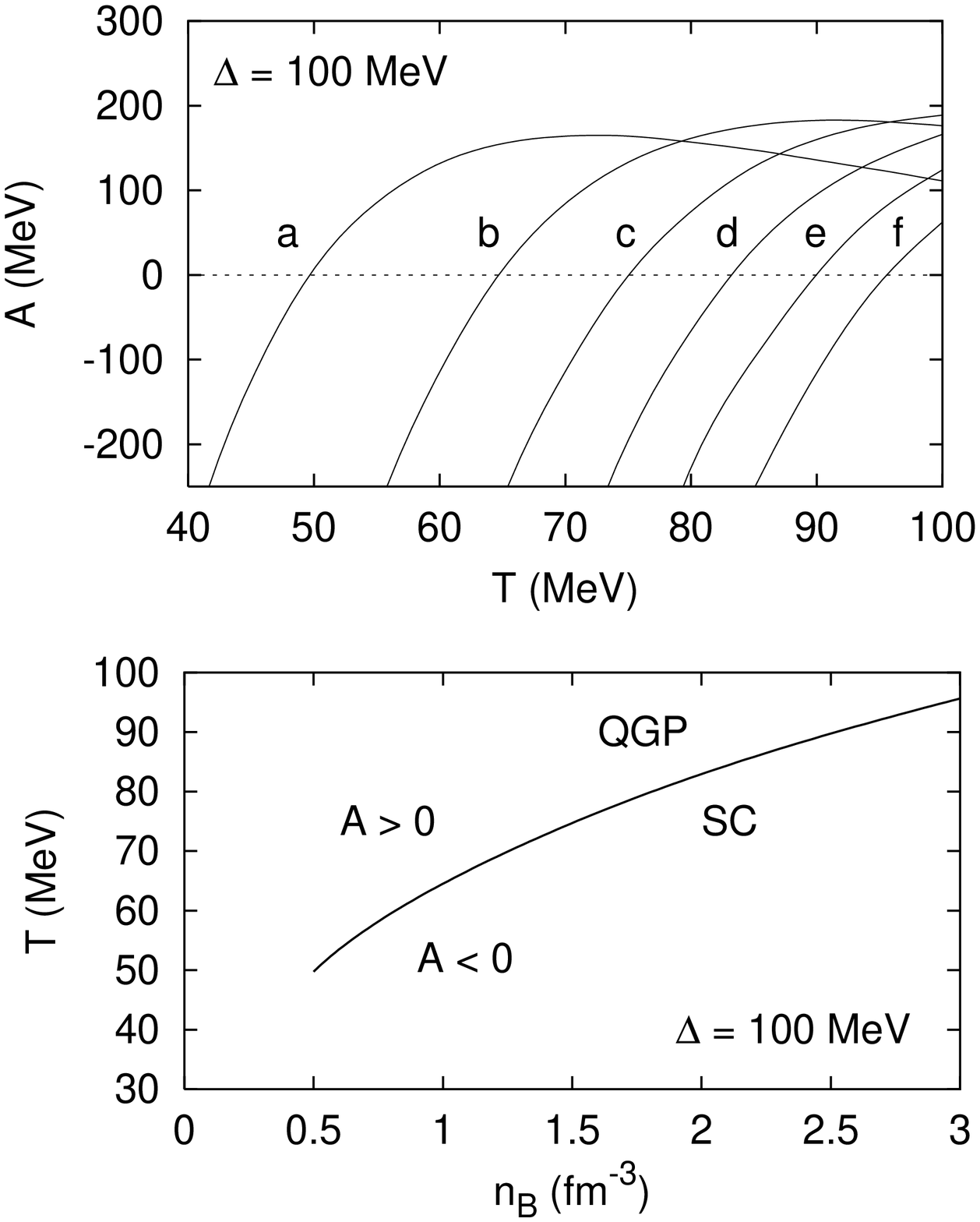}
\caption{}
\end{center}
\end{figure}

\newpage
\begin{figure}
\begin{center}
\epsfxsize=6in
\epsfbox{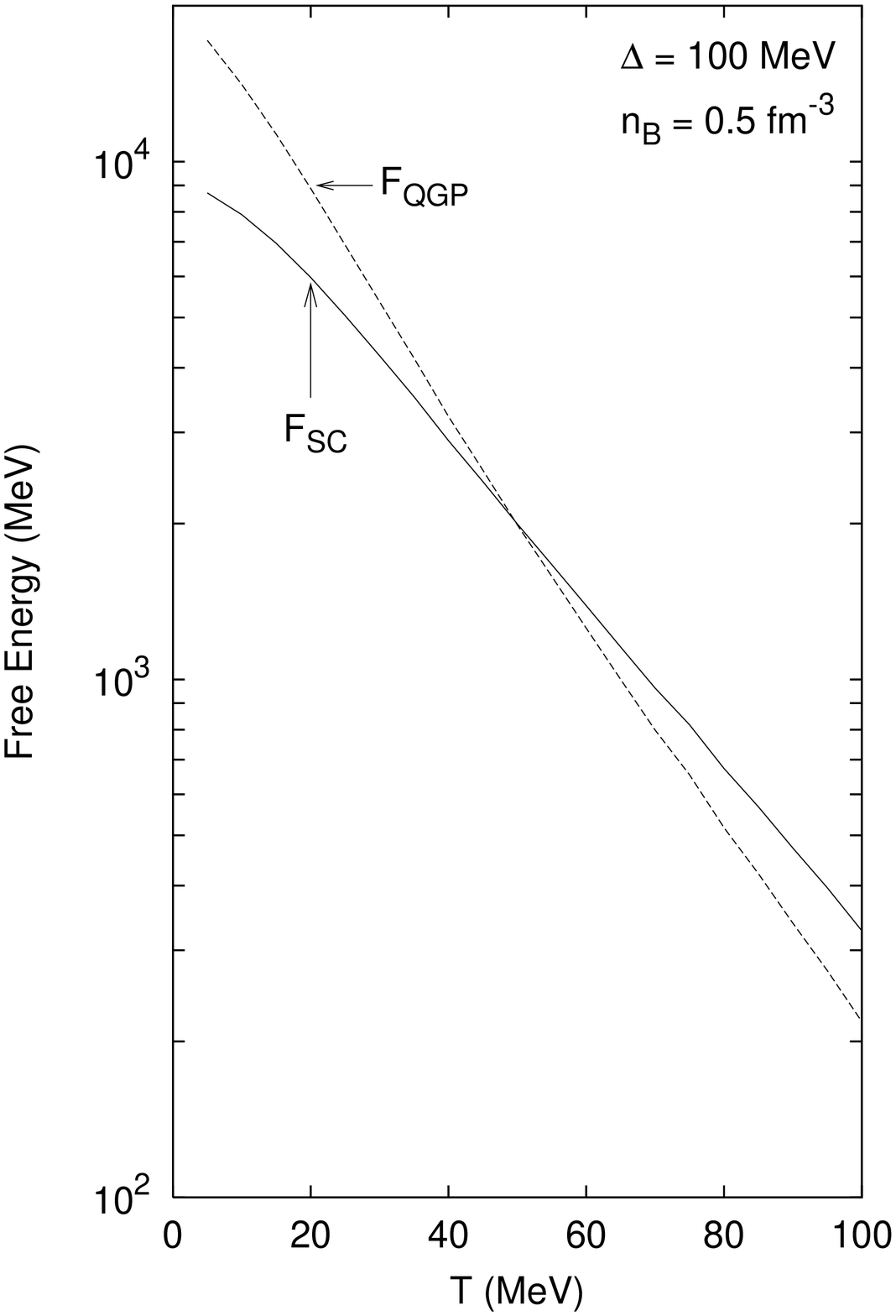}
\caption{}
\end{center}
\end{figure}

\newpage
\begin{figure}
\begin{center}
\epsfxsize=7in
\epsfbox{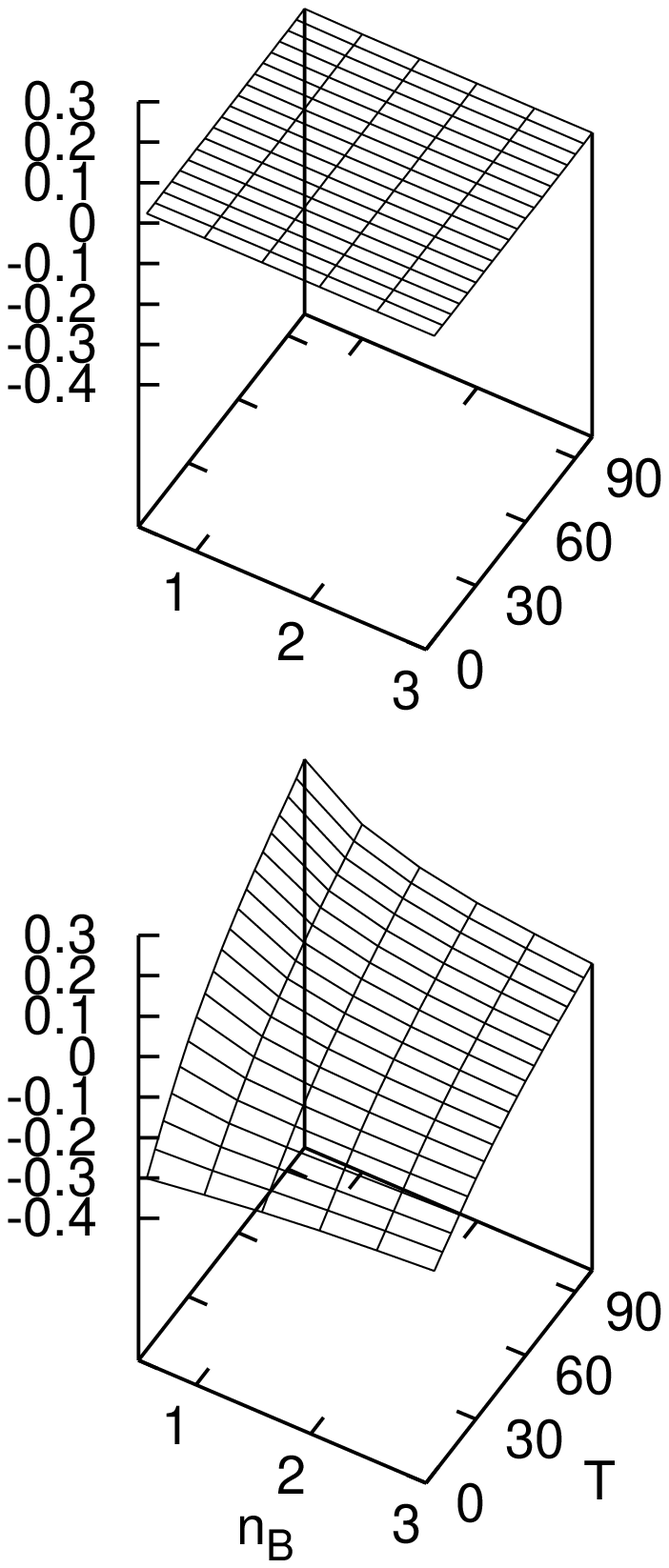}
\caption{}
\end{center}
\end{figure}

\end{document}